# The intrinsic "sense" of stochastic differential equations

Dietrich Ryter

RyterDM@gawnet.ch

Midartweg 3   CH-4500 Solothurn  Switzerland

Phone   +4132 621 13 07

A change of the variables (establishing a normal form of the forward and backward operator) neutralizes the integration sense, and the inverse singles out the Stratonovich sense, by means of the Itô formula. This excludes a free choice of the integration sense and, in particular, the Itô sense of the SDEs.

## I. Introduction

As it is well-known [1-5], the SDEs are to be understood as integral equations, and the solution is constructed by increments in $dt$. The "evaluation point" in $[t, t+dt]$ is specified by $\alpha$ ($0 \leq \alpha \leq 1$), for example at the beginning ($\alpha = 0$, Itô) or in the middle ($\alpha = 1/2$, Stratonovich). The sum of the increments may depend on $\alpha$ when $dt \to 0$; then $\alpha$ defines the "sense" of an SDE and belongs to its specification. Until now it was understood that $\alpha$ is a free parameter, to be chosen by convenience or by properties of the specific models.

A dependence on $\alpha$ does not exist when the coupling with the noise (thereby the diffusion) is constant. It will now be shown that a constant diffusion can typically be arranged by a transformation of the variables, which does *not* involve $\alpha$ (it establishes a normal form of the Fokker-Planck operator). The inverse transformation then singles out the value $\alpha = 1/2$ (Stratonovich), by applying the Itô formula to the induced map of the processes. The transformation in both directions can modify a given drift, when it does not transform as a tensor.

The transformation to the normal form has an inverse when the rank of the diffusion matrix $\underline{D}(\vec{x})$ is constant (a smooth dependence on $\vec{x}$ is assumed). It only requires quadratures, which need not even be carried out for proving the crucial statement.

The background information will briefly be summarized, particularly in the Appendix A, where the increment with any $\alpha$ is rederived.

The essential phenomena already show up in one dimension. In that case the Appendix B is trivial, and the § 3.2 reduces to its first statement; the § 4.1 shows the main findings in the simplest way.

## II. The SDE and conditional increments

The continuous Markov process $\vec{X}(t)$ is supposed to obey the SDE

$$dX^i = a^i(\vec{X})\,dt + b^{ik}(\vec{X})\,dW_k\,|_\alpha \quad \text{or} \quad d\vec{X} = \vec{a}(\vec{X})\,dt + \underline{B}(\vec{X})\,d\vec{W}\,|_\alpha \tag{2.1}$$

with smooth functions $a^i(\vec{x})$, $b^{ik}(\vec{x})$. As usual, (2.1) denotes an integral equation, with the second term involving $\alpha$. The Wiener processes $W_k(t)$ are independent and obey $<W_k(t) - W_k(0)> = 0$ and $<[W_k(t) - W_k(0)]^2> = t$. In the Appendix A it will be shown that for given $\vec{X}(t) = \vec{x}$ and $dt \geq 0$

$$\vec{X}(t+dt) - \vec{x} = \vec{a}(\vec{x})\,dt + \underline{B}(\vec{x})\,d\vec{W} + \alpha\,\vec{a}_{Sp}(\vec{x})\,dt + o(dt) \quad, \tag{2.2}$$

where $d\vec{W} = \vec{W}(t+dt) - \vec{W}(t)$, and with the "noise-induced" or "spurious" drift

$$a^i{}_{Sp}(\vec{x}) := b^{ik}{}_{,m}(\vec{x})\,b^{mk}(\vec{x}) \quad. \tag{2.3}$$

The time evolution of the probability density $w(\vec{x},t)$ of $\vec{X}(t)$ is determined by the forward (Fokker-Planck) equation [1-5]. Its "drift" is given by the expectation of (2.2)

$$<\vec{X}(t+dt) - \vec{x}> = [\vec{a}(\vec{x}) + \alpha\,\vec{a}_{Sp}(\vec{x})]\,dt + o(dt) \quad, \tag{2.4}$$

more precisely by

$$\vec{a}(\vec{x}) + \alpha\,\vec{a}_{Sp}(\vec{x}) \quad, \tag{2.5}$$

and by the "diffusion matrix"

$$\underline{D}(\vec{x}) = \underline{B}(\vec{x})\,\underline{B}^T(\vec{x}) \quad. \tag{2.6}$$

The explicit FPE reads

$$w_{,t} = [-(a^i + \alpha\,a^i{}_{Sp})\,w + (1/2)(D^{ik}w)_{,k}]_{,i} \quad. \tag{2.7}$$

The diffusion matrix (2.6) plays a key role. It is obviously symmetric and nonnegative. A constant $\underline{B}$ entails a constant $\underline{D}$. The converse clearly holds in one dimension. In higher dimensions $\underline{D}$ only determines $\underline{B}\underline{O}$ with some matrix $\underline{O}(\vec{x})$, for which $\underline{O}\underline{O}^T$ is unity, i.e. with any orthogonal $\underline{O}$ when $\det \underline{O} = 1$ ($\det \underline{O} = -1$ is also admitted). This amounts to replacing the vector Wiener process $\vec{W}$ by $\underline{O}\vec{W}$, which is stochastically equivalent [1]. In this sense a constant $\underline{D}$ also entails a constant $\underline{B}$.

In the Appendix B it will further be shown that

$$a^i{}_{Sp} := b^{ik}{}_{,j} b^{jk} = D^{ik}{}_{,k}/2 \;, \tag{2.8}$$

which is evident for a diagonal $\underline{B}(\vec{x})$, thus e.g. in one dimension.

The noise contribution in the FPE is thus completely described by $\underline{D}(\vec{x})$ and $\alpha$.

### III. Changing variables

3.1 *Tensor properties*

The variables $\vec{x}$ may be considered as coordinates in the variable space. They are now supposed to be replaced by $\vec{z}$, with a smooth and invertible transform $\vec{z}(\vec{x})$, while the Wiener process $\vec{W}(t)$ is unchanged. Since $\vec{x}$ is intrinsically contravariant, it follows (in the leading order $O(\sqrt{dt})$) that $\underline{B}(\vec{x})d\vec{W}$ is a contravariant vector (mind the lowercase argument $\vec{x}$). The dyadic product $\underline{B}d\vec{W}(\underline{B}d\vec{W})^T$ equals $\underline{B}\underline{B}^T dt$ by $d\vec{W}(d\vec{W})^T = \underline{I}dt$, and this shows that $\underline{B}\underline{B}^T = \underline{D}$ is a twice contravariant tensor. For an independent proof see [5]. The transformation of the drift $\vec{a}$ is not simple and will be considered at the end.

3.2 *Establishing a constant $\underline{D}(\vec{x})$*

In one dimension $D(x)$ becomes 1 in a new variable $z(x)$ given by $\delta z = [D(x)]^{-1/2}\delta x$, since $<(dX)>^2 = D(x)dt$ in $O(dt)$; this is confirmed by the tensor property of $\underline{D}(\vec{x})$. In higher dimensions a symmetric matrix $\underline{D}(\vec{x})$ can first be diagonalized by a field of orthogonal matrices $\underline{O}(\vec{x})$. These rotate the local coordinate axes into the eigenvectors of $\underline{D}(\vec{x})$, which are tangent to an orthogonal net $\Lambda$ of curves. New coordinates $\vec{y}$, given by $\delta\vec{y} = \underline{O}\,\delta\vec{x}$, run along these curves. The elements of the diagonal $\underline{O}^T \underline{D}\,\underline{O} := \underline{D}_d$ are the eigenvalues $\lambda_i(\vec{x})$ of $\underline{D}(\vec{x})$. The rank of $\underline{D}(\vec{x})$ is supposed to be the same for each $\vec{x}$. This means that $\lambda_i > 0$ on some curves of $\Lambda$, while $\lambda_i \equiv 0$ on the others. In the first case the rescaling of $y^i$ by

$$\delta z^i := (D_d^{ii})^{-1/2}\,\delta y^i = \lambda_i^{-1/2}\,\delta y^i \tag{3.1}$$

yields $D_c^{ii} = 1$, as in one dimension. For $\lambda_i = 0 = D_d^{ii}$ one may set $z^i = y^i$.

The constant $\underline{D}_c$ is diagonal, with elements 1 or 0. The new variables $\vec{z}$ are thus obtained in two steps: (1) by determining the eigenvalues and eigenvectors of $\underline{D}(\vec{x})$ at each $\vec{x}$, and (2) for $\lambda_i > 0$ by integrating

$$\delta z^i := \lambda_i^{-1/2}\,\delta s \qquad (\delta s \text{ being the line element}) \tag{3.2}$$

along the respective curve of $\Lambda$, while $z^i = y^i$ when $\lambda_i = 0$.

*Remarks:*

(i) In two dimensions a general nondegenerate diffusion operator was cast into a normal form by use of Beltrami equations [6]. These are fulfilled by (3.1) and (3.2), and the operator becomes the Laplacian.

(ii) The method is easily extended to more general operators, with (symmetric) coefficient matrices $\underline{D}(\vec{x})$ having eigenvalues of both signs (provided that the number of positive, zero and negative ones does not vary with $\vec{x}$); it is sufficient to take the absolute value of $\lambda_i$ in the rescaling (3.1). This establishes the normal form

$$D^{ik}(\vec{x})\,\partial^2/\partial x^i \partial x^k = \sum_i \kappa_i\,\partial^2/(\partial z^i)^2 \qquad \text{with} \quad \kappa_i = 1, 0, -1 \tag{3.3}$$

and includes, for example, the hyperbolic case.

The diffusion operator in the forward and backward equations becomes the Laplacian,

acting on the subspace where $\underline{D}$ is nonsingular.

It is important to note that the transformed $\vec{a}_{Sp}$ is zero by (2.8). This shows that $\vec{a}_{Sp}$ is *not* a tensor, because any tensor vanishing in one coordinate system vanishes altogether. Clearly a vanishing $\vec{a}_{Sp}$ neutralizes the integration sense, in view of (2.2).

### IV.  Results

4.1  *The simplest case*

The idea is most easily seen in one dimension and with $a(x) \equiv 0$ :

$$dX = b(X)\,dW \quad \text{with} \quad b(x) > 0 \text{ and with an unspecified } \alpha. \tag{4.1}$$

The transform $z(x)$ is given by $dz = [D(x)]^{-1/2}\,dx = [b(x)]^{-1}\,dx$ and results in $Z(t) \equiv W(t)$, since in the new variable $b^* \equiv 1$ [Recall that $b$ is a one-dimensional vector, with the transform $b^* = b(dz/dx) \equiv 1$, in accordance with $D^* = (b^*)^2 = 1$].

The inverse transformation $x(z)$, with $dx/dz = b[x(z)]$, expresses $X(t)$ by $Z(t) = W(t)$, and the Itô formula entails that

$$dX = b[x(z)]\,dW + (1/2)(db/dz)\,dt. \tag{4.2}$$

In view of

$$db(x)/dz = [db(x)/dx](dx/dz) = b'(x)b(x) \tag{4.3}$$

the result is just (2.2) with $\alpha = 1/2$ ($b'b = a_{Sp}$). The Stratonovich sense of (4.1) is thus imposed by the Itô formula.

*Remark:* $b(x) < 0$ is also admitted, but not a zero point of $b(x)$, where $z(x)$ would not have an inverse.

An *alternative* (and more elementary) approach makes use of the FPE

$$w_{,t} = -\alpha(bb'w)' + (1/2)(b^2 w)''.$$

Its equivalent in the $z$-variable is $u_{,t} = (1/2) u_{,zz}$ (by $b^* \equiv 1$) and does *not* contain $\alpha$ (see also (2.8)). The inverse transform will thus specify $\alpha$. Indeed, by observing that $u\,dz = w\,dx$ (giving $u = w\,dx/dz = bw$) and $\partial/\partial z = (dx/dz)\partial/\partial x = b\,\partial/\partial x$, it follows that $w_{,t} = (1/2)[-bb'w + (b^2 w)']'$, which implies that $\alpha = 1/2$. (That approach is less appropriate for generalizing).

These findings remind the well-known fact [1] that the Stratonovich integral (only) obeys the rules of conventional analysis; stochastic analysis and the FPE are essentially parallel.

### 4.2 *Higher dimensions*

The Itô formula can formally be obtained by taking the Taylor expansion to the second order in $dW_i$ and by $dW_i dW_k = \delta_{ik} dt$. This gives the clue for the extension of the above result to higher dimensions.

It is supposed that both $\vec{z}(\vec{x})$ and $\vec{x}(\vec{z})$ exist, see the preceding § 3.2. The following argument applies in the subspace where $\underline{D}(\vec{x})$ is nonsingular. Since $\underline{D}_c$ is unity there, one can modify the transformed $\underline{B}$ to become unity as well, according to the idea outlined after (2.7). Then the analogue of (4.2) is

$$dX^i = b^{ik}[\vec{x}(\vec{z})]dW_k + (1/2)\{\partial b^{ik}[\vec{x}(\vec{z})]/\partial z^m\}\,dW_m dW_k \tag{4.4}$$

with

$$\partial b^{ik}[\vec{x}(\vec{z})]/\partial z^m = (\partial b^{ik}/\partial x^n)(\partial x^n/\partial z^m) = (\partial b^{ik}/\partial x^n)b^{nm}\ . \tag{4.5}$$

Observing that $dW_m dW_k = \delta_{mk} dt$ leads to (2.2), (2.3) with $\alpha = 1/2$.

### 4.3 *A nonzero drift*

The drift $\vec{a}(\vec{x})$ becomes $\vec{a}*(\vec{z})$ in the new variables, and the new SDE reads

$$d\vec{Z} = \vec{a}*(\vec{Z})dt + \underline{I}_c\,d\vec{W}\ , \tag{4.6}$$

where the diagonal $\underline{I}_c$ is unity in the subspace where $\underline{D}(\vec{x})$ is nonsingular, and zero otherwise. This is unique and can be solved in principle (as well as the respective forward and backward equations). For the extension of the § 4.2 consider the increment of $\vec{X}(t)$

$$dX^i = (\partial x^i / \partial z^k) dZ^k + (1/2)(\partial^2 x^i / \partial z^k \partial z^m) dZ^k dZ^m .$$

Inserting (4.6), and proceeding as above in (4.4) and (4.5), results in

$$d\vec{X} = [\vec{a}**(\vec{x}) + (1/2)\vec{a}_{Sp}(\vec{x})] dt + \underline{B}(\vec{x}) d\vec{W} + o(dt) \qquad (4.7)$$

where $\vec{a}**(\vec{x})$ is the inverse transform of $\vec{a}*(\vec{z})$. Clearly $\vec{a}**(\vec{x}) \equiv \vec{a}(\vec{x})$ when $\vec{a}(\vec{x})$ is a tensor (so $a^{i*} = (\partial z^i / \partial x^k) a^k$). Otherwise $\vec{a}**(\vec{x})$ is a sort of projection of $\vec{a}(\vec{x})$, where possible derivatives of $\underline{D}(\vec{x})$, thus of $\underline{B}(\vec{x})$, are set equal to zero. This can be seen by the example $\vec{a}(\vec{x}) := \beta \vec{a}_{Sp}(\vec{x})$ with some constant $\beta$, which yields $\vec{a}* \equiv \vec{0}$ and thereby $\vec{a}** = \vec{0}$ (note that $\beta = -1/2$ would entail the Itô sense). A given drift is thus reduced to its tensor part.

## V. Comments

This work was incited by the idea that the number of solutions for models described by a SDE cannot depend on the choice of the coordinates. This would indeed be the case in the existing concept: starting from a system with a constant diffusion (thus with a unique solution) a nonlinear change of the variables entails an $\alpha$-dependence and thereby a continuum of solutions.

The new arguments rather discard the free choice of $\alpha$ and exhibit the general value $\alpha = 1/2$ (when $\alpha$ really matters). Conditions for that result were weak assumptions on the diffusion matrix: a constant rank and a smooth dependence on the argument $\vec{x}$.

This new finding is purely mathematical and must not be confused with attempts to determine $\alpha$ by external arguments (as in the "Itô or Stratonovich dilemma [7]). It rather parallels the approach by Wong and Zakai [8], but does not invoke any extra approximations or limiting procedures.

As a byproduct, a variable transform was specified, which establishes a standard form for quite general linear differential operators of the second order.

*Outlook:*

In [9] it is shown that multiplicative noise excludes the Markov property of the solution $\vec{X}(t)$. This contradicts with [1,2] and requires an extension of the existing theory. The further finding that $\alpha = 1$ admits an approximate Markov property may favor that case for practical applications.

**Appendix A**

The integral equation (2.1) is to be solved in $[t, t+dt]$, with $\vec{X}(t) = \vec{x}$. The exact increment $\Delta \vec{X}$ obeys

$$\Delta X^i (dt) = \int_t^{t+dt} a^i[\vec{x} + \Delta \vec{X}(\tau)] d\tau + \int_t^{t+dt} b^{ik}[\vec{x} + \Delta \vec{X}(\tau)] dW_k(\tau) \ . \tag{A.1}$$

For small enough $dt$ the first integral yields $a^i(\vec{x}) dt$, and the second one can be solved explicitly. To this end it is sufficient to expand $b^{ik}$ to the first order, which results in

$b^{ik}(\vec{x}) W_k(dt) + b^{ik}{}_{,m}(\vec{x}) \int_t^{t+dt} \Delta X^m(\tau) dW_k(\tau)$ (with $W_k(0) = 0$ since only the increments matter). The first term is the leading part of $O(\sqrt{dt})$, and successive approximation amounts to insert it into the integral, which results in

$$\int_t^{t+dt} \Delta X^m(\tau) dW_k(\tau) = b^{mn}(\vec{x}) \int_0^{dt} W_n(\tau) dW_k(\tau) \ .$$

The last integral involves $\alpha$. For $k = n$ it is well-known to yield $[W^2(dt) + (2\alpha - 1) dt]/2$, with the expectation $\alpha\, dt$ and with the $\alpha$-independent variance $(dt)^2/2$. For small enough $dt$ this allows to replace the integral by the nonrandom value $\alpha\, dt$. Since for $k \neq n$ the expectation is zero, the result is $\Delta X^i(dt) \approx b^{ik}(x)W_k(dt) + a^i{}_{Sp}(\vec{x})\alpha\, dt$, with the "noise-induced" or "spurious" drift

$$a^i{}_{Sp}(\vec{x}) := b^{ik}{}_{,m}(\vec{x})b^{mk}(\vec{x}) \ . \tag{A.2}$$

This yields the explicit result

$$\Delta X^i = [a^i(\vec{x}) + \alpha\, a^i{}_{Sp}(\vec{x})]dt + b^{ik}(\vec{x})W_k(dt) + o(dt) \tag{A.3}$$

and thereby the equivalent Itô form of (2.1)

$$dX^i = [a^i(\vec{X}) + \alpha\, a^i{}_{Sp}(\vec{X})]\, dt + b^{ik}(\vec{X})\, dW_k|_{(\alpha=0)} \tag{A.4}$$

for each $\alpha$.

**Appendix B**

The spurious drift $\vec{a}_{Sp}$ can always be expressed in terms of the diffusion matrix $\underline{D}(\vec{x})$. For a diagonal matrix $\underline{B}$ (of the elements $b^{ik}$) - thus in one dimension - it is obvious that

$$b^{ik}{}_{,j}\, b^{jk} = D^{ik}{}_{,k}/2 \ , \tag{B.1}$$

and for a symmetric $\underline{B}$ the same follows by diagonalizing $\underline{B}$. Each asymmetric $\underline{B}$ can be symmetrized on substituting $\vec{W}(t)$ by an equivalent $\vec{W}^*(t)$ given by $d\vec{W} := \underline{O}\, d\vec{W}^*$: With $\underline{B}^* := \underline{B}\,\underline{O}$ this entails $\underline{B}\, d\vec{W} = \underline{B}^*\, d\vec{W}^*$. When $\underline{B}$ is square, one can find a $\underline{O}$ which yields a symmetric $\underline{B}^*$ by which (B.1) holds again; a rectangular $\underline{B}$ can be completed by zeros. This shows that (B.1) holds in general (but only by stochastic equivalence when $\underline{B}$ is not symmetric):

$$a^i{}_{Sp} := b^{ik}{}_{,j}\, b^{jk} = D^{ik}{}_{,k}/2 \ . \tag{B.2}$$

.
*Acknowledgment*

The author wishes to thank H. Spohn for valuable comments on an earlier version.